\documentclass{article}

\usepackage{a4wide}
\usepackage{amsmath,amssymb,amscd,amsfonts}
\usepackage{amsthm}

\usepackage{mathrsfs,latexsym}

\newcommand{\RR}{{\mathbb R}}

\newcommand{\NN}{{\mathbb N}}

\newcommand{\ZZ}{{\mathbb Z}}

\newcommand{\CoinX}[1]{C_0^\infty({#1})}

\newtheorem{Thm}{Theorem}[section]

\newtheorem{Lem}[Thm]{Lemma}
\newtheorem{Prop}[Thm]{Proposition}
\newtheorem{Cor}[Thm]{Corollary}

\numberwithin{equation}{section}

\newcommand*\tnorm[1]{\left|\mspace{-1mu}\left|\mspace{-1mu}\left|#1\right|\mspace{-1mu}\right|\mspace{-1mu}\right| }

\DeclareMathOperator{\Ric}{Ric}

\newcommand{\supp}{{\rm supp}\,}
\newcommand\g{\gamma}
\renewcommand\th{\theta}
\renewcommand\S{\Sigma}

\newcommand\beq{\begin{equation}}
\newcommand\eeq{\end{equation}}
\newcommand\ben{\begin{enumerate}}
\newcommand\een{\end{enumerate}}
\newcommand\bit{\begin{itemize}}
\newcommand\eit{\end{itemize}}

\newcounter{mnotecount}[section]

\begin{document}
\renewcommand{\thefootnote}{\fnsymbol{footnote}}
\begin{center}
{ \Large \bf Singularity theorems from weakened energy conditions}
\\[20pt]
{\large  Christopher J.\ Fewster${}^{1}$\footnote{E-mail: chris.fewster@york.ac.uk}
 {\rm and}   Gregory J. Galloway${}^{2}$\footnote{E-mail: galloway@math.miami.edu}}
\\[20pt]  
                 ${}^1$ Department of Mathematics,
                 University of York, 
                 Heslington,
                 York YO10 5DD, United Kingdom
                 \\[10pt]
                 ${}^2$\, Department of Mathematics, University of Miami, Coral Gables, FL, U.S.A.
		 \\[18pt]
\today
\end{center}
${}$\\[10pt]
{\small {\bf Abstract. }     
We establish analogues of the Hawking and Penrose singularity theorems based on (a) averaged energy conditions with exponential damping;
(b) conditions on local stress-energy averages inspired by the Quantum Energy Inequalities satisfied 
by a number of quantum field theories. As particular applications, we establish singularity theorems for the Einstein equations coupled to a classical scalar field, which violates the strong energy condition, and the nonminimally coupled scalar field, which also violates the null energy condition. 
}
${}$

\renewcommand{\thefootnote}{\arabic{footnote}}
%

\section{Introduction}

The singularity theorems proved by Penrose and Hawking in the mid-1960's \cite{Penrose1965, Hawking1966:i, HawPen1970}
represented a major advance over previous work, which was restricted to situations of high symmetry and particular matter models. At the technical level, this advance was achieved by significant development
of the geometrical apparatus; at the physical level, however, the key idea was to replace particular
matter models by generic energy conditions, which drive geodesic congruences to focal points. 
While many classical matter models  obey
the classical energy conditions, such as perfect fluids and the electromagnetic field, there are exceptions. For example, the Klein--Gordon field with nonzero mass fails to satisfy the strong energy condition (SEC; the hypothesis for the Hawking and Hawking--Penrose theorems if the cosmological constant vanishes) while the
nonminimally coupled Klein--Gordon field also violates the null energy condition (NEC; the hypothesis of
the Penrose theorem). See~\cite{BarVis2002} for a critical assessment of the status of classical energy conditions and~\cite{Senovilla1998} for a review of singularity theorems and commentary on energy conditions in that context. 

One of the main limitations of the energy conditions is that they are incompatible with quantum field
theory, as has long been known~\cite{EGJ}. Examples of states with locally negative expectation values for the energy density are easily constructed \cite{DaviesFulling1977} and in a large class of quantum field theories the energy density at any given point is unbounded from below as the quantum state varies (see~\cite{Fews05} and references therein). Moreover, recent results in two-dimensional conformal field theory show
that individual measurements of weighted spacetime averages of energy density in the vacuum state yield negative values with probabilities approaching $90\%$ in some cases \cite{FewsterFordRoman2010}. 

For these reasons a number of authors have investigated singularity theorems under weakened energy conditions. Examples include 
\cite{Tipler1978, Borde87, Roman1988,WaldYurtsever}, in which various averaged energy conditions are considered, particularly the averaged null energy condition (ANEC), i.e., 
\[
\int_\gamma T_{ab}\g'^a \g'^b \,d\lambda\ge 0 \,,
\]
where $\gamma$ is a complete (or sometimes future- or past-complete) null geodesic in an affine parameterisation. [Via the Einstein
equations, these conditions translate immediately to conditions on the Ricci tensor.] The definition of the integral requires some care; it is usually interpreted as a $\liminf$ of the integrals over finite regions (as in \cite{Roman1988}) or regularised by inserting a mollifying factor $f(\lambda/\lambda_0)^2$ in the integrand (for $f$ in a suitable function class) and taking the $\liminf$ as $\lambda_0\to\infty$ \cite{WaldYurtsever}. On the other hand, a different approach (see 7.21--22 of~\cite{Penrose1972} and \cite{Borde1994}) is to suppose a lower bound on $R_{ab}t^a t^b$ for all unit timelike vectors and sufficient initial contraction for focussing to occur. 

In this paper, we will establish singularity theorems under conditions that combine ideas
from these two approaches 
but also extend them to allow for situations in which the integrand becomes
unboundedly negative. Roughly speaking, we replace the ANEC integral by the requirement
that for a future-complete null geodesic $\gamma:[0,\infty)\to M$,
\[
\int_\gamma e^{-c\lambda} T_{ab}\g'^a \g'^b \,d\lambda -\frac{c}{2}
\]
should be finite for some $c>0$ (with the integral understood as above)
and dominate the initial expansion of suitable future-complete null congruences
in spacetime $M$.
This leads to a generalization of the Penrose theorem, which is stated precisely
as Theorem~\ref{thm:sing2}; a similar generalization of Hawking's theorem is
given as Theorem~\ref{thm:sing1}. In Section~\ref{sect:discussion}, we apply these results to prove a Hawking singularity theorem for the massive minimally coupled Klein--Gordon field (even though it can violate the SEC), at least provided that the field amplitude is bounded or, more generally, of exponential type. Under similar conditions on the amplitude of the field and its derivatives, we also prove a Penrose singularity theorem for the nonminimally coupled Klein--Gordon field (even though it can violate the NEC). 
Our results are based on
criteria, derived in Section~\ref{sect:Riccati}, for the nonexistence of global solutions to Riccati equations
that generalize arguments of~\cite{Galloway} by inclusion of the exponential as an integrating factor (one
could equally consider other functions in this context). This idea is used to generalize
arguments of~\cite{WaldYurtsever} in Section~\ref{sect:WaldYurtsever} and the relevant
singularity theorems are proved in Section~\ref{sect:singularity}. 

In Section~\ref{sect:QEIbounds}, we consider how exponentially damped energy 
conditions of the type required can be derived from bounds on local energy averages similar to those satisfied by quantum fields, known as Quantum Energy Inequalities (QEIs) or Quantum Inequalities (QIs). Inequalities of this type, which
express limitations on the magnitude and duration of violations of the classical energy conditions, were first
mooted by Ford~\cite{Ford1978} and have since been established in a number of
quantum field theory models (see \cite{Fews05,Roma04} for reviews). 
In four-dimensional Minkowski space, for example,
the energy density of a free real scalar field obeys the bound
\[
\int_{-\infty}^\infty \langle T_{ab}\g'^a\g'^b\rangle_\psi f(t)^2\,dt \ge - \frac{1}{16\pi^2}\int_{-\infty}^\infty f''(t)^2\,dt
\]
for any smooth compactly supported real-valued function $f$ and any Hadamard state
$\psi$~\cite{FewEv98,Fews00}.\footnote{Units with $\hbar=c=G=1$ are used throughout. 
Our geometrical conventions are those of \cite{HE}.} Here $\gamma$ is any complete timelike geodesic in a proper time parameterisation. Under rescaling of $f$, the bound scales as the inverse fourth power of the sampling timescale $\tau$; this is consistent with the unboundedness below
of the energy density at points (in the limit $\tau\to 0$) and also gives an averaged weak energy condition in the limit $\tau\to\infty$. Similar results are known for other free fields. 

Although analogous results hold in curved spacetimes, the bounds are more complicated in form.
For example, the bound given in \cite{FewSmi2008} involves local Fourier transforms of
distributions formed from the first few terms in the expansion of the Hadamard parametrix,
and is valid for averaging within suitable domains that typically have compact closure. However, 
it is expected on physical grounds~\cite{FordRoman_worm1996} (and borne out in concrete
examples) that averages over timescales smaller than local curvature length scales
are constrained by bounds taking the same form as in Minkowski space. In the bounds of \cite{FewSmi2008}, for example, the first term in the Hadamard expansion will dominate the others on sufficiently small scales, and becomes well-approximated by the expression in Minkowski space. This motivates the study of energy conditions in which local averages of the energy density or other stress-energy
quantities are bounded from below in terms of $L^2$-norms of derivatives of the averaging function, perhaps with constraints on the support or after the subtraction of some reference function (as in the so-called {\em difference QEIs}). 
In Section~\ref{sect:QEIbounds} we will derive 
(variants of) the exponentially damped energy conditions from local energy conditions of this type. This prepares the way for a more detailed study, which would have to quantify the scales over which the QEI bounds can be replaced by these simpler estimates. 
We also emphasize that, while our local conditions are motivated by QEIs, they do not
coincide directly with them; see Section~\ref{sect:discussion} for discussion on this point. 

\section{Criteria for nonexistence of global solutions to the Riccati equation}
\label{sect:Riccati} 

We begin with a slight modification of an argument given in \cite[Lemma 3]{Galloway}.

\begin{Lem} \label{lem:IVP1}
Consider the initial value problem
\begin{align}
\dot{z} &= \frac{z^2}{q} + p \,,\notag\\
z(0) &= z_0 \label{eq:IVP1}
\end{align}
where $q(t)$ and $p(t)$ are continuous on $[0,\infty)$, and $q(t)>0$ on $[0,\infty)$.
If 
\[
\int_0^\infty \frac{dt}{q(t)} = +\infty \qquad\text{and}\qquad 
\liminf_{T\to +\infty}\int_0^T p(t) \,dt> -z_0
\]
then \eqref{eq:IVP1} has no solution on $[0,\infty)$. 
\end{Lem}
\proof Suppose there is a solution $z(t)$ on $[0,\infty)$.
By hypothesis, there exists $t_1\ge 0$ such that
\[
\int_0^t p(t')\,dt' > -z_0 
\]
for all $t\in [t_1,\infty)$. Integrating the differential equation,
\begin{equation}\label{eq:zineq}
z(t) = \int_0^t \frac{z(t')^2}{q(t')}\,dt' + \int_0^t p(t')\,dt' + z_0 > \int_0^t \frac{z(t')^2}{q(t')}\,dt' 
\end{equation}
for $t\ge t_1$. Introducing $R(t) = \int_0^t z(t')^2/q(t') \,dt'$, we see that $R$ is nonnegative and
obeys the differential inequality
\[
\dot{R} = \frac{z^2}{q} > \frac{R^2}{q}
\]
for $t\ge t_1$. Accordingly, we have $R(t)>0$ for all $t>t_1$. Fixing any $t_2>t_1$, we now have
\begin{equation}\label{eq:Rinvineq}
\frac{1}{R(t_2)} \ge \frac{1}{R(t_2)} - \frac{1}{R(t)} = \int_{t_2}^t \frac{\dot{R}}{R^2} dt> \int_{t_2}^t \frac{dt}{q} 
\end{equation}
for $t>t_2$.
However, the right-hand side is unbounded as $t\to\infty$ and we obtain a contradiction.
\qed

Here, `no solution' means that we have $z(t)\to +\infty$ as $t\to t_*^-<\infty$. Note that 
\[
z(t) \ge z_0 + \int_0^t p(t')\,dt'
\]
for all $t$ for which the solution exists, so $z(t)\to-\infty$ cannot occur at finite time. (It is
easily shown that divergence is the only way in which the solution can break down at finite times.)

As a digression from our main development, we note the following consequence.
\begin{Cor}\label{cor:IVP1} 
If, in Lemma~\ref{lem:IVP1}, the integral condition on $p$ is replaced by 
\[
z_0 + \inf_{T\ge 0} \int_0^T p(t)\,dt = \alpha>0 \,,
\]
then \eqref{eq:IVP1} has no solution on $[0,\tau]$, where $\tau$ is the unique solution 
to 
\[
\int_0^\tau \frac{dt'}{q(t')} = \frac{2}{\alpha}\,.
\]
In particular, this applies if the negative part $p_-(t)=\min\{0,p(t)\}$ is
integrable and $z_0 + \int_0^\infty p_-(t)dt = \alpha>0$.
\end{Cor}
\proof Following the proof of the Lemma, we may take $t_1=0$ and deduce from
the hypothesis and Eq.~\eqref{eq:zineq} 
that $z(t)\ge \alpha$ for all
$t\in[0,\infty)$ and hence that 
\[
R(t_2) \ge \alpha^2\int_0^{t_2} \frac{dt'}{q(t')}
\]
for any $t_2>0$. The inequality~\eqref{eq:Rinvineq} now asserts that
\[
\frac{1}{\alpha^2} > \left(\int_0^{t_2} \frac{dt'}{q(t')}\right) \left(\int_{t_2}^t \frac{dt'}{q(t')}\right)
\]
for all $0<t_2\le t$ and the intermediate value theorem allows us to find $t_2$ such that
both factors on the right-hand side are equal to $\frac{1}{2}\int_0^t q(t')^{-1}dt'$, 
establishing the required result. Finally, if $p_-$ is integrable, it is clear that $\int_0^T p(t)dt
\ge \int_0^\infty p_-(t)dt$ for all $T$.
\qed

\medskip
{\noindent\em \underline{Remark.}} In the standard singularity theorems, $q(t)\equiv n-1$ for timelike 
geodesic congruences and $q(t)\equiv n-2$ for null geodesic congruences, 
where $n$ is the spacetime dimension. The upper bound on the parameter
time until blow-up is therefore $\tau=2(n-1)/\alpha$ or $2(n-2)/\alpha$
respectively. Tighter bounds can be obtained in this case, by a more involved argument~\cite{Paeng2009}. 
\medskip

We now weaken the conditions on $p$; for simplicity we also fix $q$ to be a constant.
\begin{Lem} \label{lem:IVP2}
Consider the initial value problem
\begin{align}
\dot{z} &= \frac{z^2}{s} + r \,,\notag\\
z(0) &= z_0 \label{eq:IVP2}
\end{align}
where $r(t)$ is continuous on $[0,\infty)$, and $s>0$ is constant.
If there exists $c\ge 0$ such that
\[
z_0 - \frac{c}{2} + \liminf_{T\to +\infty} \int_0^T  e^{-2ct/s}r(t)\,dt  >0 
\]
then \eqref{eq:IVP2} has no solution on $[0,\infty)$. 
\end{Lem}
\proof Again, assume the contrary. Then $y(t)=(z(t)-c)e^{-2ct/s}$ solves
\begin{align}
\dot{y} &= \frac{y^2}{se^{-2ct/s}} + e^{-2ct/s}(r(t)+c^2/s)\notag \\
y(0) &= z_0-c \label{eq:IVP3}
\end{align}
on $[0,\infty)$. This equation is of the form \eqref{eq:IVP1}, with $q(t)=se^{-2ct/s}$ and $p(t) = e^{-2ct/s}(r(t)+c^2/s)$. 
Clearly, $\int_0^\infty dt/q(t)=\infty$; as we also have
\begin{align*}
\liminf_{T\to +\infty} \int_0^T e^{-2ct/s}(r(t)+c^2/s) \,dt &\ge
\liminf_{T\to+\infty} \int_0^T  e^{-2ct/s} r(t)\,dt + \liminf_{T\to +\infty}
\int_0^T  e^{-2ct/s} c^2/s\,dt \\ &= 
\frac{c}{2} + \liminf_{T\to+\infty} \int_0^T  e^{-2ct/s} r(t)\,dt   \\ &> c-z_0 = - y(0)
\end{align*}
by the hypothesis, Lemma~\ref{lem:IVP1} entails that \eqref{eq:IVP3} has no solution on $[0,\infty)$
and we have obtained a contradiction. \qed

For example, suppose that $r(t)\ge -A e^{Bt}$ for $A, B>0$. Then for $c>Bs/2$ we have
\[
-\frac{c}{2} + \liminf_{T\to +\infty} \int_0^T e^{-2ct/s} r(t)  \,dt
\ge -\frac{c}{2} - \frac{As}{2c-Bs}\,,
\]
the right-hand side of which has a maximum of $-(\sqrt{As}+Bs/4)$ at $c = \sqrt{As} + Bs/2$. 
Accordingly, \eqref{eq:IVP2} has no solution on $[0,\infty)$ if 
\begin{equation}\label{eq:exp_bd}
z_0 > \sqrt{As}+\frac{Bs}{4}\,.
\end{equation}
As an explicit example, consider $r(t) = -e^{2t}$, $s=1$ (i.e., $A=1$, $B=2$). 
The above inequality establishes nonexistence of global solutions for $z_0>3/2$; in fact, the 
equation can be solved in terms of modified Bessel functions as
\[
z(t) = \frac{d}{dt} \log \left( [ I_0(e^t)K_1(1)+K_0(e^t)I_1(1) ]- [ I_0(e^t)K_0(1)- K_0(e^t)I_0(1) ]z_0
\right)
\]
and it is easy to see that $z(t)$ has a singularity at finite time if and only if
\[
z_0 > \inf_{t\ge 0} \frac{  I_0(e^t)K_1(1)+ K_0(e^t)I_1(1) }{ I_0(e^t)K_0(1)- K_0(e^t)I_0(1) }
= \frac{K_1(1)}{K_0(1)} = 1.429...  \,,
\]
which is consistent with the estimate provided by Lemma~\ref{lem:IVP2}. 

We also have the following consequence.
\begin{Cor} \label{cor:IVP2}
If, in Lemma~\ref{lem:IVP2}, the integral condition on $r$ is replaced by
\[
z_0 -c + \inf_{T\ge 0} \int_0^T e^{-2ct/s} (r(t)+c^2/s)\,dt = \alpha>0\,,
\]
then \eqref{eq:IVP2} has no solution on $[0,\tau]$, where 
\[
\tau = \frac{s}{2c}\log(1+4c/\alpha) \,.
\]
In particular, this applies if the negative part $r_-(t)=\min\{0,r(t)\}$ is
such that $r_-(t)e^{-2ct/s}$ is integrable and $z_0 -c + \int_0^\infty e^{-2ct/s} r_-(t)dt=\alpha>0$. 
\end{Cor}
\proof Following the proof of Lemma~\ref{lem:IVP2}, the hypothesis implies that
\[
y(0) + \inf_{T\ge 0} \int_0^T p(t)\,dt = \alpha>0 
\]
and we may apply Corollary~\ref{cor:IVP1} to find that there is no solution 
to Eq.~\eqref{eq:IVP2} on $[0,\tau]$,
where $\tau$ solves
\[
\frac{2}{\alpha} =\int_0^\tau \frac{dt'}{q(t')} = \frac{e^{2c\tau/s}-1}{2c}\,,
\]
proving the first statement. The second statement follows from the 
elementary bound $\int_0^T e^{-2ct/s}(r(t)+c^2/s) \,dt
\ge \int_0^\infty e^{-2ct/s}r_-(t)\,dt$ for all $T\in [0,\infty)$.  \qed

For cases in which $\alpha\gg c$, the bound is approximately $2s/\alpha$, which is the
same as in the remark following Cor.~\ref{cor:IVP1}.

\section{A generalization of the Wald--Yurtsever argument}\label{sect:WaldYurtsever}

Wald and Yurtsever~\cite{WaldYurtsever} consider hypotheses on weighted averages of the forcing term in the Riccati eqution (see the Lemma in that reference). Here, we generalize this further, after the pattern of Lemma~\ref{lem:IVP2} above. Fix a real-valued compactly supported function $g\in\CoinX{\RR}$ with $g(t)=1$ on $[0,1]$, and $g$ nonincreasing on $\RR^+$.
\begin{Lem} \label{lem:WY_variant}
Suppose that a solution to~\eqref{eq:IVP2} exists on (all of) $t\in [0,\infty)$. If $r(t)$ is nonconstant
then 
\[
\limsup_{\tau\to\infty} \int_0^\infty e^{-2ct/s} r(t) g(t/\tau)^2 \,dt <   \frac{c}{2}-z_0 
\]
for all $c\ge 0$. 
If $r(t)\equiv r$ is constant, the same conclusion holds provided $c\neq z_0$ or $c^2/s+r\neq 0$. 
\end{Lem}
{\noindent\em \underline{Remarks.}} 
\ben
\item As the $\liminf$ of any function is always less than its $\limsup$ we may rather
trivially replace $\limsup$ by $\liminf$ in the statement of this result.  
\item The $c=0$ case essentially corresponds to the result in \cite{WaldYurtsever}. 
\een
\proof 
If a solution to~\eqref{eq:IVP2}  exists for all $t$, then we know that \eqref{eq:IVP3} has a solution on $[0,\infty)$. Rearranging, and integrating against $g(t/\tau)^2$,
\begin{equation}\label{eq:int1}
I_c(\tau):= \int_0^\infty e^{-2ct/s} r(t) g(t/\tau)^2\,dt =\int_0^\infty \dot{y}(t) g(t/\tau)^2\,dt -\frac{c^2}{s}\int_0^\infty  e^{-2ct/s} g(t/\tau)^2\,dt    - R_c(\tau)  \,,
\end{equation}
where, as before, $y(t)=(z(t)-c)e^{-2ct/s}$ obeys Eq.~\eqref{eq:IVP3} and
\[
R_c(\tau)  =  \frac{1}{s} \int_0^\infty y(t)^2 e^{2ct/s} g(t/\tau)^2\,dt  
\]
increases (not necessarily strictly) as $\tau$ increases, and is strictly positive for all $\tau>0$ unless 
$y\equiv 0$, which can happen only if $c=z_0$ and $r(t)\equiv -c^2/s$.
Integrating \eqref{eq:int1} by parts,
\begin{equation}\label{eq:int2}
I_c(\tau)= - y(0)  - \frac{2}{\tau} \int_0^\infty y(t) g(t/\tau)\dot{g}(t/\tau) \,dt
-\frac{c^2}{s}\int_0^\infty  e^{-2ct/s} g(t/\tau)^2 \,dt 
 - R_c(\tau) \,.
\end{equation}
As $y$ is continuous, both $e^{ct/s}y(t)g(t/\tau)$ and $e^{-ct/s}\dot{g}(t/\tau)$ are square-integrable, and the Cauchy--Schwarz inequality gives the estimate
\begin{align*}
\left|\int_0^\infty y(t) g(t/\tau)\dot{g}(t/\tau)\,dt  \right|
& \le \left(\frac{1}{s}\int_0^\infty y^2 e^{2ct/s}g(t/\tau)^2\,dt \right)^{1/2} 
\left(s \int_0^\infty  e^{-2ct/s}\dot{g}(t/\tau)^2 \,dt \right)^{1/2} \\
& \le C \sqrt{\tau R_c(\tau)} e^{-c\tau/s} \,,
\end{align*}
where $C>0$ is defined by 
\[
C^2=   s\int_0^\infty \dot{g}(t)^2\,dt 
\]
and we have used the fact that $\dot{g}(t/\tau)=0$ in $[0,\tau]$. 
We may also estimate 
\[
\frac{c}{2}(1-e^{-2c\tau/s})\le \frac{c^2}{s}\int_0^\infty  e^{-2ct/s}g(t/\tau)^2\,dt   \le \frac{c}{2} \,.
\]
Putting these estimates in \eqref{eq:int2}, we now have
\[
I_c(\tau) \le   - y(0)   -\frac{c}{2}(1-e^{-2c\tau/s})
- R_{c}(\tau) + 2Ce^{-c\tau/s} \sqrt{\frac{R_{c}(\tau)}{\tau}} 
\]
and, estimating 
\[
2Ce^{-c\tau/s} \sqrt{\frac{R_{c}(\tau)}{\tau}} \le \frac{R_c(\tau)}{2} + 2C^2 \frac{e^{-2c\tau/s}}{\tau}
\]
by the AM--GM inequality, 
we arrive at 
\[
I_c(\tau) \le   - y(0)   -\frac{c}{2}(1-e^{-2c\tau/s})
- \frac{R_{c}(\tau)}{2} + 2C^2\frac{e^{-2c\tau/s}}{\tau} \,.
\]
If either (a) $r(t)$ is nonconstant, or (b) $r(t)\equiv r$ but $z_0\neq c$ or $c^2/s+r\neq 0$, then
$y$ cannot be identically zero. It follows that $R_{c}(\tau)$ is strictly positive and increasing on $(0,\infty)$. Thus there exists $\tau_0>0$ such that 
\[
I_c(\tau)  <  - y(0)   -\frac{c}{2} = \frac{c}{2}-z_0
\]
for all $\tau>\tau_0$, which proves the required result. \qed

\section{QEI-inspired hypotheses}\label{sect:QEIbounds}

Now let us consider a QEI-inspired condition: we suppose that the forcing function $r(t)$ 
is constrained by inequalities of the form
\begin{equation}\label{eq:EC1}
\int_{-\infty}^\infty  (r(t)-r_0(t))  f(t)^2 dt \ge - \tnorm{f}^2
\end{equation}
for all real-valued $f\in\CoinX{\RR}$, where $r_0(t)$ is a fixed continuous real-valued function and $\tnorm{\cdot}$ is any Sobolev (semi-)norm of the form
\[
\tnorm{f}^2 = \sum_{\ell=0}^L Q_\ell \|f^{(\ell)}\|^2 \,,
\]
where $Q_\ell\in [0,\infty)$ are constants of the appropriate dimensions, $L\in\NN$ and $\|\cdot\|$ is the $L^2$-norm.

In the analogy with QEIs, $r_0$ would be obtained from a reference state.  As already mentioned,
this constraint is inspired by, rather than exactly coinciding with, the known QEI results. 
See Section~\ref{sect:discussion} for more discussion.

Let $g$ be as above and $c\ge 0$. Choose a smooth real-valued function $h\in C^\infty(\RR)$, such that 
$\supp h \subset [-\tau_0,\infty)$ and $h(t)= e^{-ct/s}$ on $[0,\infty)$. Then for each $\tau>0$,
\[
f_\tau(t) =\begin{cases} e^{-ct/s}g(t/\tau) & t>0 \\
h(t) & t<0
\end{cases}
\]
defines a test-function $f_\tau\in \CoinX{\RR}$. If $c>0$ then the dominated convergence theorem entails
that $\tnorm{f_\tau}\to \tnorm{h}$ as $\tau\to\infty$ and we have
\[
\liminf_{\tau\to+\infty} I_c(\tau) \ge 
-\int_{-\tau_0}^0 h(t)^2 (r(t) -r_0(t))\,dt - \tnorm{h}
+\liminf_{\tau\to+\infty} \int_0^\infty e^{-2ct/s} r_0(t)g(t/\tau)^2 \,dt \,.
\]
The same holds if $c=0$, provided that the coefficient $Q_0$ vanishes in the definition
of $\tnorm{\cdot}$. Note that $\tnorm{h}$ can be written 
\[
\tnorm{h} = \sum_{\ell=0}^L  Q_\ell\left(\frac{1}{2} (c/s)^{2\ell-1} + \int_{-\tau_0}^0 |h^{(\ell)}(t)|^2\,dt\right)\,.
\]
The following result is now immediate. 
\begin{Thm}\label{thm:QEI}
Suppose $r(t)$ obeys \eqref{eq:EC1} and is nonconstant. Suppose there exist $c> 0$, $\tau_0>0$
and $h\in C^\infty(\RR)$ obeying $\supp h \subset [-\tau_0,\infty)$ and $h(t)= e^{-ct/s}$ on $[0,\infty)$,
for which
\[
z_0 - \frac{c}{2} + \liminf_{\tau\to+\infty} \int_0^\infty e^{-2ct/s} r_0(t)g(t/\tau)^2 \,dt \ge \int_{-\tau_0}^0 h(t)^2 (r(t) -r_0(t))\,dt +  \tnorm{h}^2 \,.
\]
Then~\eqref{eq:IVP2} has no solution on all of $[0,\infty)$. If $r(t)\equiv r$ obeys \eqref{eq:EC1},
the same conclusion holds provided $c\neq z_0$ or $c^2/s+r\neq 0$. If $Q_0=0$,
the hypothesis on $c$ can be relaxed to $c\ge 0$.
\end{Thm}
\proof By Lemma~\ref{lem:WY_variant}, if there is a solution to~\eqref{eq:IVP2} on $[0,\infty)$ then
\begin{align*}
0 &> z_0-\frac{c}{2} +  \liminf_{\tau\to\infty} I_c(\tau)  \\
& \ge z_0-\frac{c}{2}
+\liminf_{\tau\to+\infty} \int_0^\infty e^{-2ct/s} r_0(t)g(t/\tau)^2 \,dt 
 -\int_{-\tau_0}^0 h(t)^2 (r(t) -r_0(t))\,dt - \tnorm{h}^2  \\
&\ge 0
\end{align*}
by hypothesis, which is a contradiction. \qed

The above theorem tells us that a focal point will occur if there is sufficient initial contraction. Note that the dependence on the actual matter distribution $r(t)$ is purely through its values
in $[-\tau_0,0]$, i.e., before the contraction $z_0$ is measured. At first sight it is strange that {\em positive} values of $r(t)-r_0$ require a larger value of $z_0$: the reason is simply that the averaged nature of the energy condition means that large positive energies in the present allow large negative values in the future. 
The magnitude of initial contraction required is otherwise determined by the reference $r_0$ and the function $h$. 

In some QEIs, the support of the sampling function is constrained, for example, 
to be small relative to local curvature length scales. 
For simplicity, let us suppose that there is some $\tau_0$ such that \eqref{eq:EC1} is valid for all $f$ with support in an interval of length at most $2\tau_0$. (This would correspond to conditions in which 
curvatures remain bounded; if they do not, this might be regarded as an indication of singular behaviour, albeit not in the sense of geodesic incompleteness.) We also restrict to a particular seminorm which would be most relevant to QEIs in four dimensions.

To discuss averages over longer timescales we will use a partition of unity. To this end, we will
say that $\psi\in\CoinX{\RR}$ is a {\em bump function} if it is nonnegative,  supported in $[0,3/2]$ and obeys $\psi(3/4+x)=\psi(3/4-x)$, $\psi(x)^2+ \psi(1/2-x)^2 = 1$
for $x\in[0,3/4]$; in particular, this gives $\psi=1$ on $[1/2,1]$. Then the
functions $\psi(x-n)^2$ ($n\in\ZZ$) form a partition of unity on $\RR$. Note that this partition involves
the squares of the underlying functions. The main work is to show that this gives usable bounds
on the exponentially damped averages appearing in Lemma~\ref{lem:WY_variant}. 

\begin{Prop}\label{prop:partition} 
Let $r_0$ be a fixed continuous function and suppose that 
$r(t)$ is nonconstant and obeys 
\begin{equation}
\label{eq:EC2}
\int_{-\infty}^\infty  (r(t)-r_0(t))  f(t)^2 dt \ge -Q \|f''\|^2
\end{equation}
for any $f\in\CoinX{(0,\infty)}$ supported in an interval of length at most $2\tau_0>0$. Let $\psi$ be a bump function. Then for any $c>0$ we have
\begin{align}
\liminf_{\tau\to +\infty}  \int_0^\infty e^{-2ct/s} r(t)g(t/\tau)^2\,dt &\ge 
\liminf_{\tau\to +\infty}  \int_0^\infty e^{-2ct/s} r_0(t)g(t/\tau)^2\,dt 
-\int_{-\tau_0}^0 \!\!\! (r(t)-r_0(t))  h(t)^2\,dt  \notag\\
&\qquad\qquad -Q \left( \|h''\|^2 + \frac{1}{2}\left(\frac{c}{s}\right)^3
+ \frac{\|\psi''\|^2}{\tau_0^3(1-e^{-2c\tau_0/s})}\right) \,. \label{eq:EC3}
\end{align}
for all real-valued $h\in\CoinX{(-\tau_0,\tau_0/2)}$ with
$h(t) = e^{-ct/s}\psi(1/2-t/\tau_0)$ on $t\ge 0$. The last term may be 
replaced by
\[
 -Q \left( \int_{-\tau_0}^0 |h''(t)|^2\,dt + \left(\frac{c}{s}\right)^3
+\frac{\|\psi''\|^2}{\tau_0^3} \left(\frac{1}{2}+\frac{1}{(1-e^{-2c\tau_0/s})}\right)\right) \,.
\]
\end{Prop}

We defer the proof to the end of this section. The following consequence is immediate, by the same reasoning as in Theorem~\ref{thm:QEI}. 
\begin{Thm}\label{thm:QEI2}
Let $r_0$ be a fixed continuous function and $\psi$ be a bump function.
Suppose $r(t)$ is nonconstant and obeys~\eqref{eq:EC2}
for any $f\in\CoinX{(0,\infty)}$ supported in an interval of length at most $2\tau_0>0$. If, for some $c> 0$ and $h$ obeying the conditions of Prop.~\ref{prop:partition}, we have
\begin{equation}
z_0   \ge  \frac{c}{2} - (\textrm{RHS of~\eqref{eq:EC3}}) \label{eq:QEI2_hyp}
\end{equation}
then~\eqref{eq:IVP2} has no solution on all of $[0,\infty)$. If $r(t)\equiv r$ obeys \eqref{eq:EC2},
the same conclusion holds provided $c\neq z_0$ or $c^2/s+r\neq 0$. 
\end{Thm}
\medskip
{\noindent\underline{\em Remark.}}  
It is possible to find bump functions $\psi$ with $\|\psi''\|^2\sim 331$. To see this, 
consider the ansatz
\[
\psi(x) = \begin{cases}  \sin(\theta(x)) &  x\le 1/4 \\ \cos(\theta(1/2-x)) & 1/4<x\le 3/4
\\ \psi(3/2-x) & x>3/4 \end{cases}
\]
for $\theta\in C^\infty(\RR)$, $\theta\equiv 0$ on $\RR^-$, $\theta\ge 0$
and $\theta(1/4) = \pi/4$, and $\theta^{(2k)}(1/4)=0$ for all $k\in\NN$, which
ensure that $\psi$ is smooth at $x=1/4$. 
After a calculation, one finds that
\[
\|\psi''\|^2 = 2\int_0^{1/4} \left(\theta'(x)^4 + \theta''(x)^2\right)\,dt \,,
\]
which corresponds to the Euler--Lagrange equation
\begin{equation}\label{eq:EL}
\theta''''(x) - 6\theta'(x)^2 \theta''(x) = 0
\end{equation}
together with boundary conditions $\theta(0)=\theta'(0)=0$, $\theta(1/4) = \pi/4$, $\theta''(1/4) = 0$.
Numerical solution of this ODE in Maple~14 gives the value $\|\psi''\|^2 = 330.97$. This does not 
extend to give a smooth solution to our original problem as $\theta''(0+)=44.56$; however, 
by considering a mollified function $\theta_\epsilon(x) = \theta(x)f(x/\epsilon)$ with $f\in C^\infty(\RR)$,
$f(x)=0$ on $(-\infty,0]$, $f(x)\equiv 1$ on $[1,\infty)$, one may show that
$\|\psi_\epsilon''\|\to\|\psi''\|$ as $\epsilon\to 0^+$ [one uses estimates $\theta(x)=O(x^2)$, 
$\theta'(x)=O(x)$, $\theta''(x)=O(1)$ near $x=0$].

\proof[Proof of Proposition~\ref{prop:partition}] The idea of the proof is to
use a partition of unity to decompose the integral 
\[
J_c(\tau) = \int_{0}^\infty \!\! e^{-2ct/s} (r(t)-r_0(t))  g(t/\tau)^2\, dt \,,
\]
thereby obtaining a sum of integrals, each of which can
be bounded below using~\eqref{eq:EC2} (only finitely many terms
appear at any fixed value of $\tau$). The sum of the resulting lower bounds must be
controlled as $\tau\to\infty$. 

We start from the identity
\begin{align}
\left(\frac{d^2}{dt^2} e^{-ct/s}\varphi(t)\right)^2 &= e^{-2ct/s}\left(\frac{c}{s}\right)^4 \left\{
\varphi(t)^2 - 2\frac{s}{c}\frac{d}{dt}(\varphi^2) + 2\left(\frac{s}{c}\right)^2 \varphi'(t)^2
+\left(\frac{s}{c}\right)^2\frac{d^2}{dt^2}(\varphi^2) \notag \right.\\
&\qquad\qquad\left.
-2\left(\frac{s}{c}\right)^3 \frac{d}{dt}((\varphi')^2) +\left(\frac{s}{c}\right)^4\varphi''(t)^2
\right\}\,.
\label{eq:basic_ident}
\end{align}
For real-valued $\varphi\in\CoinX{(0,\infty)}$, we
may integrate by parts and discard a nonpositive term to find
\begin{align}
\int_0^\infty \left(\frac{d^2}{dt^2} e^{-ct/s}\varphi(t)\right)^2 dt &= \left(\frac{c}{s}\right)^4 \int_0^\infty
e^{-2ct/s} \left(\varphi(t)^2 -2\left(\frac{s}{c}\right)^2 \varphi'(t)^2 + 
\left(\frac{s}{c}\right)^4 \varphi''(t)^2\right)\,dt \notag\\
&\le \int_0^\infty
e^{-2ct/s} \left(\left(\frac{c}{s}\right)^4\varphi(t)^2  + 
\varphi''(t)^2\right)\,dt \,.\label{eq:useful_calc}
\end{align}

If, additionally, the support of $\varphi$ is contained in an interval of length at most
$2\tau_0$, we may apply~\eqref{eq:EC2} with  $f(t) =e^{-ct/s} g(t/\tau)\varphi(t)$ and $g$ chosen as above, to give
\begin{align}
\lefteqn{\int_{0}^\infty \!\! e^{-2ct/s} (r(t)-r_0(t))  g(t/\tau)^2\varphi(t)^2 \,dt}   \notag \\ 
&\qquad\qquad\ge -
 Q\int_{0}^\infty  \!\! e^{-2ct/s} \left(\frac{c^4}{s^4} 
g(t/\tau)^2\varphi(t)^2  + \left(\frac{d^2}{dt^2}g(t/\tau)\varphi(t)\right)^2\right) \,
dt,
\label{eq:EC2phig} 
\end{align}
where we have estimated $\|f''\|^2$ using~\eqref{eq:useful_calc}, with 
$\varphi$ replaced by $g(t/\tau)\varphi(t)$. This estimate will be used for each term appearing in the partition of unity, to which we now turn.

Let $\varphi(t)=\psi(t/\tau_0)$, where $\psi$ is the chosen bump function.  Defining $\varphi_n(t) = \varphi(t-nt_0)$,  the $\varphi_n^2$ form a partition of unity so that $\sum_{n=0}^\infty
\varphi_n(t)^2 \equiv 1$ on $t>\tau_0/2$ and at most two $\varphi_n(t)$ are nonzero
at each $t$; each $\varphi_n$ has support diameter of $3\tau_0/2$. We may now decompose the integral $J_c(\tau)$ using this partition, estimating each term
using~\eqref{eq:EC2phig}. This gives
\begin{align*}
J_c(\tau) &= \int_0^{\tau_0/2}\!\!  e^{-2ct/s} (r(t)-r_0(t))  g(t/\tau)^2(1-\varphi(t)^2)\, dt
+ \sum_{n=0}^\infty \int_0^\infty \!\! e^{-2ct/s} (r(t)-r_0(t))  g(t/\tau)^2\varphi_n(t)^2\, dt\\
&\ge \int_0^{\tau_0/2}\!\!  e^{-2ct/s} (r(t)-r_0(t))  g(t/\tau)^2(1-\varphi(t)^2)\, dt   - Q \sum_{n=0}^\infty S_n(\tau) \,,
\end{align*}
where
\[
S_n(\tau) =\int_{0}^\infty  e^{-2ct/s} \left(\left(\frac{c}{s}\right)^4 g(t/\tau)^2\varphi_n(t)^2  + \left(\frac{d^2}{dt^2}g(t/\tau)\varphi_n(t)\right)^2\right)
\,dt \,.
\]

Our task is now to control the sum $\sum_{n=0}^\infty S_n(\tau)$ in the limit
$\tau\to\infty$. Now, for $\tau>(n+3/2)\tau_0$, $g(t/\tau)\equiv 1$ on the 
support of $\varphi_n$, and we obtain
\begin{align*}
S_n(\tau)&=\int_{0}^\infty  e^{-2ct/s} \left(\left(\frac{c}{s}\right)^4 \varphi_n(t)^2  + \varphi_n''(t)^2\right)
\,dt \\
&\le\left(\frac{c}{s}\right)^4 \int_{0}^\infty  e^{-2ct/s}  \varphi_n(t)^2  dt
+ e^{-2cn\tau_0/s}\|\varphi''\|^2\,,
\end{align*}
where the exponential factor in the second term arises on account of the support
properties of $\varphi_n$ and we use $\|\varphi_n''\|=\|\varphi''\|$. Provided the
limit and sum can be exchanged (which will be justified below) we then have
\[
\lim_{\tau\to\infty} \sum_{n=0}^\infty S_n(\tau) 
\le \sum_{n=0}^\infty 
\left(\frac{c}{s}\right)^4 \int_{0}^\infty  e^{-2ct/s}  \varphi_n(t)^2  dt
+ \|\varphi''\|^2 \sum_{n=0}^\infty e^{-2cn\tau_0/s}\,.
\]
The first sum can be evaluated using the partition of unity property and the second
is elementary, so
\begin{align*}
\lim_{\tau\to\infty} \sum_{n=0}^\infty S_n(\tau) 
&\le \left(\frac{c}{s}\right)^4\int_{0}^{\tau_0/2} e^{-2ct/s} (\varphi(t)^2-1) dt + 
 \left(\frac{c}{s}\right)^4  \int_{0}^{\infty} e^{-2ct/s}dt 
+ \frac{\|\varphi''\|^2}{1-e^{-2c\tau_0/s}} \\
&\le 
\frac{1}{2}\left(\frac{c}{s}\right)^3
+ \frac{\|\varphi''\|^2}{1-e^{-2c\tau_0/s}}\,,
\end{align*}
where we discard the nonpositive contribution from $\varphi^2-1$ in the last step.

Assembling the results so far, we have 
\[
\liminf_{\tau\to +\infty} J_c(\tau) \ge
\int_0^{\tau_0/2}  e^{-2ct/s} (r(t)-r_0(t))  (1-\varphi(t)^2) dt   - Q\left( 
\frac{1}{2}\left(\frac{c}{s}\right)^3
+ \frac{\|\psi''\|^2}{\tau_0^3(1-e^{-2c\tau_0/s})}\right) \,.
\]
Taking $h$ as specified in the hypotheses, we note that $h(t)^2 = e^{-2ct/s}(1-\varphi(t)^2)$
on $t\ge 0$. Applying Eq.~\eqref{eq:EC2} with $h$ in place of $f$, we obtain
\[
\liminf_{\tau\to +\infty} J_c(\tau) \ge
-\int_{-\tau_0}^0   (r(t)-r_0(t))  h(t)^2\, dt   - Q\left( \|h''\|^2+
\frac{1}{2}\left(\frac{c}{s}\right)^3
+ \frac{\|\psi''\|^2}{\tau_0^3(1-e^{-2c\tau_0/s})}\right) \,.
\]
Finally, because 
\[
\liminf_{\tau\to +\infty} I_c(\tau)  \ge \liminf_{\tau\to+\infty}J_c(\tau) + \liminf_{\tau\to+\infty}\int_0^\infty e^{-2ct/s} r_0(t)\,dt\,,
\]
we obtain the the required result~\eqref{eq:EC3}. The alternative form of the
bound arises if we estimate $\|h''\|^2$ by using \eqref{eq:basic_ident} applied to
$\varphi(t)=\psi(1/2-t/\tau_0)$ and integrating by parts on $[0,\infty)$ to obtain \eqref{eq:useful_calc} (the boundary terms at $t=0$ cancel). Using the properties
of $\psi$ we find
\[
\|h''\|^2 \le \int_{-\tau_0}^0 |h''(t)|^2\,dt + \frac{1}{2}\left(\frac{c}{s}\right)^3 + 
\frac{1}{2\tau_0^3}\|\psi''\|^2
\]
which yields the alternative bound.

It remains to justify the exchange of limit and sum used above, for
which we use a dominated convergence argument. We split $S_n(\tau)$
into the sum of two terms. For the first, we note that
\[
\int_{0}^\infty  e^{-2ct/s}g(t/\tau)^2\varphi_n(t)^2
\,dt  \le e^{-2cn\tau_0/s}\|\varphi\|^2
\]
because $\supp\varphi_n\subset [n\tau_0,\infty)$. The second is regarded as the
square of an $L^2$ norm, which can be estimated using the Leibniz rule and triangle inequalities as 
\[
\left[\int_{0}^\infty  e^{-2ct/s}\left(\frac{d^2}{dt^2}g(t/\tau)\varphi_n(t)\right)^2
\,dt\right]^{1/2}  \le \frac{C_n}{\tau^2} +
\frac{C_n'}{\tau} +  C_n'' \,,
\]
where $C_n$, $C_n'$ and $C_n''$ are the $L^2$ norms of $e^{-ct/s}g''(t/\tau)\varphi_n(t)$, $2e^{-ct/s}g'(t/\tau)\varphi_n'(t)$ and  $e^{-ct/s}g(t/\tau)\varphi_n''(t)$ respectively. Setting $C=\max\{1,4\|g'\|^2_\infty,\|g''\|_\infty^2\}$ and again using the support properties of $\varphi_n$, we have
\[
\int_{0}^\infty  e^{-2ct/s}\left(\frac{d^2}{dt^2}g(t/\tau)\varphi_n(t)\right)^2
\,dt  \le C \tau_0 e^{-2cn \tau_0/s}\left(\frac{\|\psi\|}{\tau^2}+
 \frac{\|\psi'\|}{\tau\tau_0}+\frac{\|\psi''\|}{\tau_0^2}\right)^2\,.
\]
Accordingly,  there exists $K>0$ such that $0\le S_n(\tau) \le K e^{-2cn\tau_0/s}$ for all $n\in\NN_0$ and $\tau>\tau_0$, where $K$ is independent of $\tau$ and $n$.
As this upper bound is evidently summable, the dominated convergence theorem
permits us to exchange the limit and sum as claimed. 
\qed

\section{Singularity theorems}\label{sect:singularity}

The results from the previous sections can be used to obtain refinements of some of the classical  singularity theorems of Hawking and Penrose \cite{HE} that allow global violations of the
classical energy conditions. 

We first consider the cosmological setting.  

\begin{Thm}\label{thm:sing1}
Let $M$ be a globally hyperbolic spacetime of dimension $n \ge 2$, and let $S$ be a smooth compact  spacelike Cauchy surface for $M$.  Suppose along each future complete unit speed timelike geodesic $\g : [0,\infty) \to M$ issuing orthogonally from  $S$,  there exists 
$c \ge 0$ such that, 
\beq\label{eq:econd1}
 \liminf_{T\to \infty} \int_0^T  e^{-2ct/(n-1)}r(t) \,dt > \th(p) +  \frac{c}{2}  \,,
\eeq
where $r(t) := {\rm Ric}\,(\g'(t),\g'(t)) = R_{ab}\g'^a\g'^b(t)$
and $\th(p)$ 
is the  {\it expansion} (i.e., mean curvature) of $S$ at $p = \g(0)$.  Then $M$ is future timelike geodesically incomplete.
\end{Thm} 

\noindent
\underline{\it Remarks.}  
\ben
\item  As usual, if one assumes that  the Einstein equations,
\beq\label{eq:einstein}
R_{ab} - \frac12 R g_{ab} = 8\pi T_{ab} 
\eeq
hold then the Ricci curvature term  can be directly related to the energy-momentum tensor $T_{ab}$, and, hence,  one should view (\ref{eq:econd1})  as an energy condition on spacetime.
\item Note that the  energy condition (\ref{eq:econd1}) is satisfied provided along each such timelike geodesic
$\g$, the condition,
\[
 \liminf_{T\to \infty} \int_0^T r(t) \, dt> \th(p) 
\]
(corresponding to $c=0$) holds.  This condition emphasizes the fact that if $S$ is mean contracting, i.e., if $\th$ is negative on $S$, then \eqref{eq:econd1} can hold even if $r(t)$ is everywhere negative.
\item Specializing further, if the {\it strong energy condition} holds,  i.e., if 
${\rm Ric}\,(X,X) = R_{ab}X^aX^b \ge 0$ for all timelike vectors $X$, and if $S$ is everywhere mean contracting then \eqref{eq:econd1} holds, and we essentially recover Hawking's cosmological singularity theorem; cf., Theorem 4 in \cite[p 272]{HE}. (For convenience we have stated Theorem \ref{thm:sing1} as a future singularity result rather than a past singularity
result.)
\een

\proof[Proof of Theorem \ref{thm:sing1}]   We construct an $S$-ray $\g$,
i.e., a future inextendible timelike geodesic $\g$ emanating from a point on $S$ that realizes the
Lorentzian distance to $S$ from each of its points, as follows.   Choose a sequence of points 
$q_n$  which extends arbitrarily far into the future of $S$.\footnote{More precisely, if $h$ is a complete background Riemannian metric on $M$, choose $q_n$ in $J^+(S)$ so that the $h$-distance from
$S$ to $q_n$ tends to infinity as $n \to \infty$.}  By properties of Cauchy surfaces, there exists  a timelike geodesic segment  $\g_n$ from
$p_n \in S$ to $q_n$ that realizes the Lorentzian distance from $S$ to $q_n$.   Since $S$ is compact, by taking a subsequence if necessary, we may assume that the sequence $p_n$ converges to a point $p \in S$.  Let $\g: [0,a) \to M$, $a  \in (0, \infty]$,  be the  future inextendible  unit speed timelike geodesic issuing orthogonally from $p \in S$.  The maximality of the 
$\g_n$'s guarantees that $\g$ is an $S$-ray (see, for example, the proof of the Sublemma in~\cite{Gal86}).

Let $\rho: J^+(S) \to \RR$ be the Lorentzian distance function from $S$,
$$
\rho(x) = d(S, x) = \sup_{y \in S} d(y,x) \,.
$$
By global hyperbolicity,  $\rho$ is continuous on $J^+(S)$, 
and will be smooth up to the focal cut locus
of $S$ (see \cite{Kemp}). Since $\g$ is an $S$-ray, there are no focal points, or focal cut points, to $S$ along $\g$, which, by the lower semi-continuity of the $S$-distance-to-cut locus function \cite{Kemp},
ensures that $\rho$ is smooth on a neighbourhood $U$ of $\g$.
On this neighbourhood
$u:= -\nabla\rho$ is  a smooth future directed, geodesic, hypersurface orthogonal unit timelike vector field such that $u = \g'$ along $\g$.   Consider the expansion scalar $\th=  {\rm div}\,u$. Along $\g$, 
$\th = \th(t)$, $t \in [0,a)$,
obeys Raychaudhuri's equation (for an irrotational timelike congruence) \cite{HE},
\beq\label{eq:ray}
\frac{d \th}{d t}  = - {\rm Ric}\,(\g',\g') - 2 \sigma^2 - \frac1{n-1} \th^2   \,,
\eeq
where $\sigma$ is the shear scalar.

We now observe that $\g$ is necessarily future incomplete. 
For suppose that $\gamma$ is future complete (i.e., suppose $a=\infty$). Setting $z = - \th$, $r ={\rm Ric}\,(\g',\g') + 2\sigma^2$, $s = n-1$ and $z(0) = - \th(p)$ in \eqref{eq:IVP2}, we see that Lemma \ref{lem:IVP2}, together with the energy condition  \eqref{eq:econd1}, implies that \eqref{eq:ray}, with $\th(0) = \th(p)$, has no solution on $[0,\infty)$, which is a contradiction.\qed

\smallskip
\noindent
\underline{\it Remarks.}  
\ben

\item Although slightly more complicated to state, Lemma \ref{lem:WY_variant} and Theorems 
\ref{thm:QEI} and \ref{thm:QEI2} provide alternative energy conditions that yield singularity theorems similar to Theorem \ref{thm:sing1}.

\item The assumption of global hyperbolicity in Theorem \ref{thm:sing1} can be relaxed.  It is sufficient to assume that $M$ admits a smooth compact acausal spacelike hypersurface $S$.
In this case one can construct an $S$-ray $\g$ contained in the future domain of dependence 
$D^+(S)$ (see especially \cite[Main Lemma]{Gal86}), and use the fact that $S$ is a Cauchy surface for the total domain of dependence $D(S)$.

\item We see from the proof that it would be sufficient for \eqref{eq:econd1} to hold on $S$-rays.
\een

\medskip

We now consider an extension of the Penrose singularity Theorem \cite[Theorem 1]{HE}   to the energy conditions being considered here.   Let $\S$ be a codimension two compact acausal spacelike submanifold in a spacetime $M$ of dimension $n \ge 3$.  Under suitable orientation assumptions, there exist two smooth nonvanishing independent null normal vector fields, $\ell_+$ and $\ell_-$,  along $\S$,  corresponding to outgoing and ingoing light-rays emanating from $\S$.  The null expansion scalars $\th_+$ and $\th_-$ on $\S$ are obtained by taking the divergence of $\ell_+$ and $\ell_-$, respectively, along $\S$,  and measure the instantaneous divergence of the outgoing and ingoing light rays emanating from $\S$. In a strong gravitational field both $\th_+$ and $\theta_-$ can be negative, in which  case $\S$ is called a trapped surface.   According to the Penrose singularity theorem, if $\S$ is a trapped surface in a spacetime $M$ having a noncompact Cauchy surface and satisfying the null energy condition, then $M$ is future null geodesically incomplete. 

The following generalizes the Penrose singularity theorem (see also \cite[Theorem 2]{GalSen}). 

\begin{Thm}\label{thm:sing2}
Let $M$ be a  spacetime of dimension $n \ge 3$ with a  noncompact Cauchy surface $S$.
Let $\S$ be a smooth compact acausal spacelike submanifold of $M$ of codimension two, with null expansion scalars $\th_{\pm}$ associated to the future directed null normal vector fields
$\ell_{\pm}$.   Suppose along each future complete affinely parameterized null geodesic
$\eta : [0,\infty) \to M$,  issuing orthogonally from $\S$ with initial tangent 
$\ell_{\pm}$,  there exists $c \ge 0$ such that, 
\beq\label{eq:econd2}
 \liminf_{T\to \infty} \int_0^T  e^{-2ct/(n-2)}r(t)\,dt> \th_{\pm}(p) +  \frac{c}{2}  \,,
\eeq
where $p = \eta(0)$ and $r(t) := {\rm Ric}\,(\eta'(t),\eta'(t)) = R_{ab}\eta'^a\eta'^b(t)$.  Then $M$ is future null geodesically incomplete.
\end{Thm} 

\noindent
{\it Remarks:}  If $\S$ is a trapped surface and $M$ obeys the null energy condition, then \eqref{eq:econd2} is satisfied for sufficiently small $c$ and we recover the
Penrose singularity theorem.  The case in which \eqref{eq:econd2} holds with $c =0$ corresponds to the codimension two case of Theorem 2 in \cite{GalSen}.  As in Theorem \ref{thm:sing1}, Lemma \ref{lem:WY_variant} and Theorems \ref{thm:QEI} and \ref{thm:QEI2} may be used to provide alternative energy conditions that yield singularity theorems similar to Theorem \ref{thm:sing2}.  Finally note that
if the Einstein equations \eqref{eq:einstein} hold,  then the Ricci curvature term  ${\rm Ric}\,(\eta'(t),\eta'(t))$ can be replaced by $8\pi T_{ab}\eta'^a\eta'^b(t)$.

\proof[Proof of Theorem \ref{thm:sing2}]  We shall be brief, as the general structure of the proof is similar to the proof of  the Penrose singularity theorem.  See \cite{HE} for further details and relevant results from causal theory.

The {\it achronal boundary} $\partial J^+(\S)$ is a $C^0$ achronal hypersurface ruled by null geodesics, i.e., for each point $q \in \partial J^+(\S) \setminus \S$ there exists a future directed 
null geodesic segment from a point in $\S$ to $q$ which is entirely contained in $\partial J^+(\S)$.  Such null geodesics are called the null generators of $\partial J^+(\S)$. 

If $\partial J^+(\S)$ were compact, then  flowing along the integral curves of a timelike vector
field on $M$  would establish a homeomorphism between $\partial J^+(\S)$ and the Cauchy surface $S$, which would contradict the noncompactness of $S$.  Thus we may assume $\partial J^+(\S)$ is noncompact.   Since $\partial J^+(\S)$ is closed but noncompact we can find a sequence of points $q_n$ in  $\partial J^+(\S)$ whose $h$-distance to $\Sigma$ (where $h$ is a complete background Riemannnian metric on $M$) tends to infinity as $n \to \infty$.  Let $\eta_n$ be a null
geodesic generator of $\partial J^+(\S)$ from $p_n \in \S$ to $q_n$; $\eta_n$ must meet $\S$ orthogonally, otherwise the achronality of $\partial J^+(\S)$ would be violated. By passing to a subsequence if necessary, we may assume without loss of generality that $p_n \to p \in \S$ and
that the initial tangents $\eta_n'(0) \to \ell_+(p)$.   

Let $\eta : [0, a) \to M$, $a \in (0,\infty]$,  be the affinely parameterized future inextendible null geodesic emanating from $p$ with initial tangent  $\ell_+(p)$; $\eta$ is contained in 
$\partial J^+(\S)$ since each $\eta_n$ is.  Since  $\partial J^+(\S)$ is achronal there can be 
no null focal points, or null focal cut points, to $\S$ along $\eta$.  It follows \cite{Kemp}  that $\eta$ is contained
in a smooth null hypersurface $H \subset \partial J^+(\S)$  generated by null geodesics emanating from $\S$ near $p$ with initial tangents given by $\ell_+$.  Let $K$ be a smooth null vector field on $H$ such that $K = \eta'$ along $\eta$, and let $\hat\th$ be the null expansion of $H$ with respect to $K$.    Along $\eta$,  $\hat\th = \hat\th(t)$, $t \in [0,a)$,
obeys Raychaudhuri's equation (for an irrotational null congruence) \cite{HE},

\beq\label{eq:ray2}
\frac{d \hat \th}{d t}  = - {\rm Ric}\,(\eta',\eta') -2 \hat\sigma^2 - \frac1{n-2}\hat \th^2   \,.
\eeq
Now one can argue just as in the proof of Theorem \ref{thm:sing1}.  If $\eta$ were future complete,
so that $a = \infty$, then   Lemma \ref{lem:IVP2} and  \eqref{eq:econd1}  would imply that
\eqref{eq:ray2}, with $\hat\th(0) = \th_+(p)$,  has no solution on $[0,\infty)$. Hence, $\eta$
must be future incomplete.\qed
%
%
%

\section{Applications and discussion}\label{sect:discussion}

We begin with two applications to the Einstein equations coupled to a real scalar field. The minimally coupled Einstein--Klein--Gordon system in $n>2$ spacetime dimensions is given by
\[
R_{ab} -\frac{1}{2}Rg_{ab} = 8\pi T_{ab}^{\textrm{min}}\,, \qquad T_{ab}^{\textrm{min}} =  \nabla_a\phi \nabla_b\phi -\frac{1}{2}g_{ab}(
\nabla^c\phi\nabla_c\phi + m^2\phi^2)\,.
\]
If $\gamma$ is a unit speed timelike geodesic then
\[
r(t) = {\rm Ric}(\gamma',\gamma') = 8\pi \left((\nabla_{\gamma'}\phi)^2 - \frac{m^2}{n-2}\phi^2\right)\,,
\]
from which we can see easily that (see, e.g., \cite{Bekenstein1975} and~\cite[p.~95]{HE}) the Klein--Gordon field fails to obey the strong energy condition and hence the standard hypotheses of the Hawking singularity theorem. However, if $\phi$ remains bounded along $\gamma$, with  $|\phi| \le \phi_*$, say, then we have
\[
-\frac{c}{2} + \int_0^T  e^{-2ct/(n-1)}r(t)\, dt \ge 
-\frac{c}{2}-\frac{K^2}{2c}
\]
for all $c,T>0$, where $K=m\phi_*\sqrt{8\pi(n-1)/(n-2)}$. The right-hand side has a maximum for $c=K$, and
we may conclude from Theorem~\ref{thm:sing1} that if $\theta< -K$ everywhere on $S$ then $M$ is future timelike geodesically incomplete. 

In fact, the same result follows from Theorem~4 of~\cite{Borde1994} (in the $n=4$ case). However, we may also obtain more general results, which both allow for exponential growth
of $\phi$ in the proper time along these geodesics and provide finer detail in the case of exponential decay. For example, if $S$ is a smooth compact spacelike Cauchy surface and the scalar field obeys a bound
\[
|\phi(p)|\le \phi_* e^{a\rho(p)/(n-1)}
\]
for $p\in J^+(S)$, where $\rho(p)$ is the Lorentzian distance from $p$ to $S$ and $a\in\RR$ is constant,
then we obtain a bound 
\[
-\frac{c}{2} + \int_0^T  e^{-2ct/(n-1)}r(t)\, dt \ge 
-\frac{c}{2}-\frac{K^2}{2(c-a)}
\]
for any $c> a$ and all $T>0$ along $S$-rays, for which $\rho(\gamma(t))=t$. Optimising over $c>0$ as before, and using Theorem~\ref{thm:sing1} (and the third remark following its proof) we then have that
$M$ is future timelike geodesically incomplete if 
\[
\theta< \begin{cases} -a/2-K & a\ge -K \\ K^2/(2a) & a<-K\end{cases}
\]
on $S$. The case $a=0$ corresponds to the result above, while the other cases indicate the additional
power of our technique.

A drawback of these results is the need to 
invoke the supremum $\phi_*$ and (if needed) the constant $a$. In principle they are implicit in the Cauchy data for the Einstein--Klein--Gordon system on $S$. One can also read this result in the following way (with $a=0$ for simplicity): 
if $S$ is mean contracting then either $M$ is future timelike geodesically incomplete or
$\phi$ exceeds $m^{-1}\sqrt{(8\pi)^{-1}(n-2)/(n-1)}\inf_S |\theta|$ in magnitude somewhere to the
future of $S$. 

We now turn to the nonminimally coupled field of mass $m\ge 0$ and coupling $\xi$, with stress-energy tensor
\[
T_{ab} = T^{\textrm{min}}_{ab} + \xi
\left(g_{ab}\square - \nabla_a\nabla_b + (R_{ab}-\frac{1}{2}Rg_{ab})\right)\phi^2\,.
\]
The presence of the Einstein tensor in this expression complicates the initial value problem for the Einstein equations with this source (see~\cite{Huebner1995} for existence and uniqueness results). In addition, 
the second derivatives allow violations of the NEC as well as the SEC, which can be exploited to
construct nonsingular cosmological solutions~\cite{Bekenstein1974}. However, 
for $\xi\in[0,1/4]$ (including the conformal coupling $\xi=\frac{1}{4}(n-2)/(n-1)$)
the theory obeys the inequality
\[
\int_\gamma T_{ab} \g'^a \g'^b f(\lambda)^2\,d\lambda \ge 
-2\xi \int_\gamma \left\{ f'(\lambda)^2 -\frac{1}{2}{\rm Ric}\,(\g',\g') f(\lambda)^2 -\left(\frac{1}{4}-\xi\right)R\g'^2 \right\}\phi^2\,d\lambda
\]
for any affinely parameterized causal geodesic $\gamma$ and smooth compactly supported, real-valued $f$ (\cite[Theorem II.1]{FewOst2006},
modulo change in conventions). If we assume that the field magnitude remains bounded, $|\phi|\le
\phi_*$ this entails (for the case of null $\gamma$) that 
\[
\int_\gamma {\rm Ric}\,(\g',\g') (1 - 8\pi\xi \phi^2) f(\lambda)^2\,d\lambda 
\ge -16\pi\xi\phi_*^2\|f'\|^2
\]
for all $f\in\CoinX{\RR}$. Provided that $\phi_*$ is
strictly less than the critical value $(8\pi\xi)^{-1/2}$ we may absorb a factor into $f$, thus obtaining
\begin{align*}
\int_\gamma {\rm Ric}\,(\g',\g') f(\lambda)^2\,d\lambda &\ge -16\pi\xi\phi_*^2
\int_{-\infty}^\infty \left(\frac{d}{d\lambda} \frac{f(\lambda)}{\sqrt{1-8\pi\xi\phi^2}}\right)^2 \,d\lambda \\
&\ge -Q\left(  \|f'\|^2  +\tilde{Q}^2\|f\|^2 \right) =:-\tnorm{f}
\end{align*}
for all real-valued $f\in\CoinX{\RR}$, where 
\[ 
Q = \frac{32\pi\xi\phi_*^2}{1-8\pi\xi\phi_*^2}\,,\qquad \tilde{Q} = \frac{8\pi\xi\phi_*\phi_*'}{1-8\pi\xi\phi_*^2}\,,
\]
and $\phi'_*$ is an upper bound on $|\phi'|$.
(We have used the simple estimate $\|(fg)'\|^2 \le 2(\|f\|^2\|g'\|_\infty^2 + \|f'\|^2\|g\|_\infty^2)$ for any $f\in\CoinX{\RR}$, $g\in C^\infty(\RR)$.) This is a bound of the form~\eqref{eq:EC1} with $r_0\equiv 0$ and $r={\rm Ric}\,(\g',\g')$. To proceed, we need the following lemma.

\begin{Lem} 
Let $\tau_0>0$, $s>0$ and choose $K>0$ so that
\[
K^2 \ge \tilde{Q}^2 + Q^{-1}{\rm Ric}\,(\g',\g')
\]
on $(-\tau_0,0]$. For any $\epsilon>0$, there exists $c>0$ and $h\in C^\infty(\RR)$ with
$\supp h \subset [-\tau_0,\infty)$,  $h(t)= e^{-ct/s}$ on $[0,\infty)$ and
\[
\frac{c}{2} + \tnorm{h} + \int_{-\tau_0}^0 h^2{\rm Ric}\,(\g',\g')\,d\lambda
\le \tilde{Q}\sqrt{Qs+ Q^2/2}+\frac{1}{2}QK\coth K\tau_0+\epsilon.
\]
\end{Lem}
\proof We take $c=\tilde{Q}s\sqrt{2Q/(Q+2s)}$. Then the left-hand side becomes
\[
\tilde{Q}\sqrt{Qs+ Q^2/2}+\int_{-\tau_0}^0 \left(Qh'^2+ [Q\tilde{Q}^2+\Ric\,(\g',\g')]h^2\right)\,d\lambda
\]
so it is sufficient to show that
\begin{equation}\label{eq:hineq}
\inf_h \int_{-\tau_0}^0 \left(h'^2+ K^2 h^2\right)\,d\lambda \le \frac{K}{2}\coth K\tau_0\,,
\end{equation}
with the infimum taken over the class of $h$ specified in the hypotheses. Treating this as a variational
problem, the Euler--Lagrange equation is $h''=K^2 h$ and applying the boundary conditions
$h(-\tau_0)=0$, $h(0)=1$, the solution is $h(\lambda)=(\sinh K\tau_0)^{-1}\sinh K(\lambda+\tau_0)$,
giving equality in \eqref{eq:hineq}. This can be approximated arbitrarily well within the given class of $h$.
\qed

Accordingly, using Theorem~\ref{thm:QEI} in place of Lemma~\ref{lem:IVP2}, we obtain an analogue of Theorem~\ref{thm:sing2} if the hypothesis~\eqref{eq:econd2} 
is replaced by the requirement
\[
\theta_\pm(p) < -\tilde{Q}\sqrt{(n-2)Q+ Q^2/2} - \frac{1}{2}QK\coth K\tau_0\,,
\]
with $K$ computed as above for the extension of the null geodesic $\eta$ to $(-\tau_0,0]$, for some
$\tau_0>0$. 

Finally, we briefly discuss the potential for applications involving quantum fields. The QEIs established for many free fields are weakened versions of the weak and dominant energy conditions, while our analogue of the Hawking singularity theorem is based on a weakened strong energy condition. In order to prove a Hawking result from QEI hypotheses, it would be necessary to add conditions on the trace of the stress-energy tensor, much as in the case of the classical scalar field discussed above. In addition, it is necessary to find estimates on the timescales for averaging over which the curved spacetime QEIs are well-approximated by Minkowski space results. Nonetheless, there seems a good prospect of obtaining results along these lines, and our results provide a proof of principle for the idea that  energy conditions based on local averages can be used to deduce singularity theorems. Turning to the Penrose-type results, a significant problem at present is that 
no locally averaged energy inequality along individual null geodesics is known, and direct analogues of the
results for timelike averaging cannot hold in dimensions higher than $2$~\cite{FewsRoma2003}. Here, it seems that the best approach is to consider averages with a degree of transverse smearing, whereupon QEIs can be proved~\cite{FewsRoma2003}. It is worth recalling that even the ANEC condition
is problematic for averaging along a complete null geodesic: the real scalar field violates ANEC in general spacetimes (see~\cite{UrbanOlum2010a} and references therein) although it holds in Minkowski space \cite{WaldYurtsever} and in general spacetimes along complete achronal geodesics with a tubular Minkowskian neighbourhood~\cite{FewOlumPfen2007}. Moreover, this problem persists even with some transverse averaging schemes~\cite{UrbanOlum2010b}.

Finally, it should be noted that a full treatment of singularity theorems in the context of quantized matter 
would require (at least) a semiclassical analysis that takes backreaction into account in a dynamical fashion,
bringing significant technical challenges -- see \cite{Pinamonti2010, EltzGott2010} for recent results in simple cosmological models and further references, and \cite{FlanWald1996} for positive results on ANEC with transverse smearing in this context.\footnote{The examples studied in~\cite{UrbanOlum2010b} are not 
solutions to the semiclassical equations.}

%
%
%
%
%

\medskip
\noindent {\em Acknowledgments:} 
We are grateful to the Centre for Analysis and Nonlinear PDEs (Maxwell 
Institute) and the ICMS in Edinburgh for financial support during the workshop 
Mathematical Relativity at which this work was initiated. It is a pleasure to thank the workshop organisers and participants for useful discussions, particularly Paul Tod and James Vickers. This work was partially supported by NSF grant DMS-0708048.

\providecommand{\newblock}{}

\end{document}